\documentclass[twocolumn,aps,preprintnumbers,prb,floatfix,showpacs,amsmath,amsfonts,amssymb,floatfix,superscriptaddress,
]{revtex4-1}

\usepackage{graphics}
\usepackage{color}

\usepackage{amssymb}
\usepackage{amsmath}
\usepackage{overpic}
\usepackage{epstopdf}

\usepackage{bm}
\usepackage{graphicx}

\setcitestyle{numbers,square}

\begin{document}

\title{Evidence for Correlated Dynamics near the Berezinskii-Kosterlitz-Thouless-like
Transition in a Highly Underdoped La$_{2-x}$Sr$_{x}$CuO$_{4}$}

\author{Zhenzhong Shi}
\affiliation{National High Magnetic Field Laboratory, Florida State University, Tallahassee, Florida 32310, USA}
\author{Xiaoyan Shi}
\affiliation{Department of Physics, University of Texas at Dallas, Richardson, TX 75080, USA}
\author{Dragana Popovi\'c}
\email{dragana@magnet.fsu.edu}
\affiliation{National High Magnetic Field Laboratory, Florida State University, Tallahassee, Florida 32310, USA}

\date{\today}

\begin{abstract}
A low-frequency resistance noise study in highly underdoped thick films of La$_{2-x}$Sr$_{x}$CuO$_{4}$ ($x=0.07$ and 0.08) reveals slow, correlated dynamics and breaking of ergodicity near the superconducting transition of the Berezinskii-Kosterlitz-Thouless type.  The observed correlated behavior is strongly suppressed by disorder.
\end{abstract}

\maketitle

\section{Introduction}
\label{intro}

The dynamics of topological defects has been of great interest in condensed matter physics \cite{Thouless-topology} and beyond \cite{cosmology-topology}.    Physical realizations include various two-dimensional (2D) systems, such as superfluid helium films \cite{Bishop1978}, liquid crystals \cite{Birgeneau1978}, superconducting (SC) films and Josephson-junction arrays \cite{review_minnhagen}, magnetic films \cite{Elmers1996}, and ultracold atoms \cite{Hadzibabic2006}.  Such 2D systems exhibit a phase transition at a critical temperature $T_{BKT}\neq 0$, driven by the Berezinskii-Kosterlitz-Thouless (BKT) mechanism \cite{Berezinskii1972,Kosterlitz1973,bkt2}.  In this picture, logarithmically interacting topological defects are bound in vortex-antivortex (V-AV) pairs in the low-$T$ phase, while in the high-$T$ phase defects unbind and single-vortex excitations proliferate.  Despite several decades of intensive research on this class of topological phase transitions \cite{jose_book}, experiments on the dynamics near the BKT transition remain scarce, even for the 2D SC transition, likely the most studied example of the BKT physics.  Meanwhile, theoretical studies have revealed the ergodicity breaking \cite{Faulkner2015} and the onset of out-of-equilibrium dynamics near $T_{BKT}$ \cite{Jelic2011}.  However, except for a recent study of thin films of NbN \cite{Koushik2013}, there have been no attempts, to the best of our knowledge, to probe the correlations in the BKT critical regime and the low-$T$ phase.  Furthermore, studies of the BKT physics in thin films of conventional superconductors, such as NbN, are complicated by the still not-well-understood relationship between film thickness, disorder, and superconductivity \cite{Ivry2014}.  Hence, alternative approaches are needed to explore correlations near the BKT transition.

Layered superconductors with weak interlayer Josephson coupling, such as underdoped cuprates, present an alternative route for the exploration of the BKT behavior \cite{blatter_review,emery_95,review_lee,benfatto_mu_prl07,Baity2016}.  We report a study of the dynamics near the SC, BKT-like transition, in which a) measurements are performed on highly underdoped, atomically smooth thick films of La$_{2-x}$Sr$_{x}$CuO$_{4}$ (LSCO), a prototypical cuprate, and b) resistance ($R$) noise spectroscopy, a relatively uncommon technique, is used to probe the dynamics on orders-of-magnitude longer time scales than those explored previously in other systems near the BKT transition.  In addition to the critical slowing down consistent with the BKT physics, we find evidence for the onset of \textit{correlated} dynamics and breaking of ergodicity.  Furthermore, the correlated behavior is strongly suppressed by disorder.
 
While 2D systems cannot have a true long-range order, the BKT theory \cite{Berezinskii1972,Kosterlitz1973,bkt2} shows that they may develop a quasi-long-range order at low $T$, such that the spatial correlations decay in a power-law fashion.  Therefore, the entire low-$T$ ($T<T_{BKT}$) phase appears to be scale invariant or critical.  The system is characterized by a finite superfluid stiffness $J_s$, as topological excitations are bound in V-AV pairs.  At $T=T_{BKT}$, thermal unbinding of V-AV pairs leads to a discontinuous drop of $J_s$, the key signature of the BKT physics.  On the other hand, in the disordered ($T> T_{BKT}$) phase, the spatial correlations decay exponentially with distance.  As $T\rightarrow T_{BKT}^{+}$, the correlation length $\xi$ diverges exponentially, and so does the corresponding time scale $\tau_{\xi}\sim\xi^{z}$ ($z=2$ is the dynamical exponent), signifying critical slowing down in analogy with other phase transitions.  $\xi$ corresponds to the average separation between free vortices [$\xi^2(T)\sim 1/n_F$; $n_F$ is the inverse density of free vortices above $T_{BKT}$], leading to $R\sim\xi^{-2}$.  However, in addition to free vortices, bound V-AV pairs with sizes smaller than $\xi$ still remain, and the system retains a superfluid stiffness on short length scales.  Therefore, at $T> T_{BKT}$, the correlations that persist on short enough length and time ($t$) scales can be probed using finite-frequency ($f$) measurements.

Finite-$f$ probes, which have generally involved measurements of the complex impedance and magnetic flux noise, have provided evidence for critical slowing down near $T_{BKT}$ (e.g. Refs. [\onlinecite{Liu2011}] and [\onlinecite{Shaw1996}], respectively), but not for correlations.  Also, in complex impedance studies, disorder-induced inhomogeneity may lead to spurious effects at low $f$ \cite{ganguly_prb15}, and magnetic flux noise is more susceptible to surface \cite{rogers1992} and geometrical \cite{Kim1999b} effects.  In contrast, $R$ noise has been a powerful probe of correlated dynamics in many systems \cite{Weissman1993,Bogdanovich2002,Jaroszynski2002,Jaroszynski2004,Lin2012}, including lightly doped, insulating LSCO \cite{Raicevic2008,Raicevic2011}. Being a bulk probe, it is less sensitive to surface and geometrical effects, and it measures $R$ directly. Near the BKT transition, however, it has been used to study correlations only in disordered thin ($\sim 3$-6~nm) films of NbN \cite{Koushik2013}, in which classical percolation or disorder was found to be also important. Therefore, a careful study of $R$ noise on a drastically different system is crucially needed to reveal the intrinsic, material-independent dynamics of the BKT transition. 

In contrast to conventional isotropic superconductors, such as NbN \cite{Mondal2011} and TiN \cite{Baturina_2012}, in which BKT physics can be observed only in ultrathin films, in bulk layered systems the relevant natural scale for 2D physics is the interlayer distance $d_c$, not the film thickness $d$ \cite{blatter_review}.  For weak interlayer Josephson coupling, as in underdoped cuprates \cite{emery_95}, the SC transition is still driven by the unbinding of V-AV pairs in each plane, in analogy with the purely 2D case \cite{shenoy_prl94,friesen_prb95,pierson_prb95,olsson_prb91,benfatto_mu_prl07,sondhi_prb09}.  When the vortex-core energy $\mu$ is large, the BKT transition is moved to a $T$ higher than the one expected for each isolated layer, resulting in an effective sample thickness larger than $d_c$ \cite{benfatto_mu_prl07}.  Indeed, a recent study of highly underdoped, 150 CuO$_2$ layers thick ($d_{c}=6.6$~\AA; $d\approx 100$~nm) films of LSCO found \cite{Baity2016} the BKT-like transition controlled by $J_s$ of 2-3 layers, and a large $\mu/\mu_{XY}\approx1.4$ ($\mu_{XY}$ is the value expected in the standard 2D $XY$ model \cite{bkt2,review_minnhagen}), in contrast to BCS superconductors in which $\mu/\mu_{XY}\simeq 0.2$ \cite{Mondal2011,yong_prb13}.  While the BKT physics is relevant below the mean-field critical temperature $T_c$, an extended regime of 2D Aslamazov-Larkin-type \cite{AL1,AL2,Varlamov_book} Gaussian SC (amplitude and phase) fluctuations was observed above $T_c$ \cite{Baity2016}.

\section{Experiment}

\begin{figure}
\includegraphics[scale=0.3]{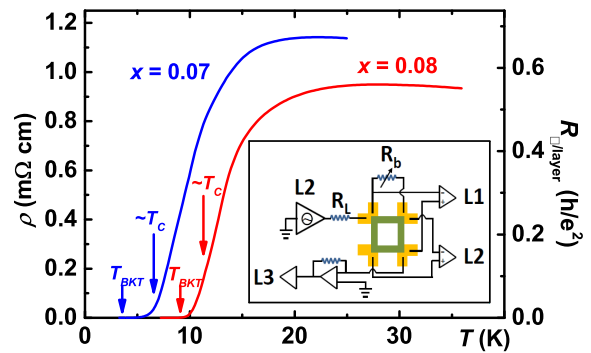}
\caption{Resistivity $\rho$ vs $T$ for $x=0.07$ and 0.08 LSCO samples ($R_{\square/\textrm{layer}}$ is $R$ per square per CuO$_2$ layer).  Arrows point at $T_{BKT}=T_{R=0}$ and $T_{c}$; $T_{c}$ was obtained from the dc transport study of the adjacent Hall bar samples on the same films \cite{Baity2016}.  Inset: A schematic of the sample bridge geometry and the measurement circuit.  Each sample has four arms (green) and eight Au (yellow) contact pads. The dimensions of each arm of the bridge are $200~\mu$m~$\times~20~\mu$m.  A constant ac excitation current $I_{exc}\approx 10~\mu$A  is obtained by connecting $R_{L}= 100$~k$\Omega$ in series with the ac ($\sim$13 Hz) voltage output of a SR 7265 lock-in amplifier (L2).  The voltage along the sample, used to measure the average $R$, and the voltage fluctuations across the bridge, $V(t)$, are measured simultaneously with lock-in amplifiers L1 and L2, respectively. $I(t)$ is measured with an Ithaco 1211 current preamplifier and another lock-in amplifier (L3).  The variable resistor $R_b$ balances the bridge.}
 \label{Fig1}
\end{figure}

We report the in-plane $R$ noise measurements on samples with an eight-contact bridge geometry (Fig.~\ref{Fig1} inset) that were patterned on the same molecular-beam-epitaxy-grown LSCO films with $x=0.07$ and 0.08 used in the dc transport study \cite{Baity2016}.  The samples become SC below $T_{BKT}= (3.5\pm 0.3)$~K and $(9.1\pm 0.5)$~K, respectively (Fig.~\ref{Fig1}), where $T_{BKT}$ is identified as the temperature at which $R=0$, consistent with the analysis in Ref.~\onlinecite{Baity2016}. 
%
%
Indeed, these values are, within error, the same as those obtained on the adjacent Hall-bar samples studied in Ref.~\onlinecite{Baity2016}.  Therefore, the values of $T_{c}$ are also expected to be comparable, i.e. $T_c\sim 6.5$~K and $\sim 11.3$~K for $x=0.07$ and 0.08, respectively.  

The time-averaged resistance $R$ and the resistance fluctuations $\Delta R(t)=R(t)-R$ were measured simultaneously \cite{Palanisami2005a,Palanisami2005} using a standard ac ($\sim 13$~Hz) technique (Fig. \ref{Fig1} inset).  Resistance fluctuations in the sample throw the bridge out of balance, resulting in voltage fluctuations $V(t)$ measured across the bridge.  The sample bridge geometry minimizes the effects of $T$ fluctuations, contact noise, and excitation source fluctuations \cite{Black1982,Alers1989,Nevins1990,Nevins1992}, as confirmed by the negligible correlations between the measured $I$ and $V$ fluctuations.  Likewise, great care was taken to ensure the measured noise was free from $T$ fluctuations ($ \leq 1$~mK) even for measurements as long as 10 hours. Most of the data were obtained in the $f=(10^{-4}-10^{-1})$~Hz bandwidth, much lower than that in the 
magnetic flux noise studies [e.g. $(10^{-1}-10^{4})$~Hz in Refs.~\onlinecite{Shaw1996,Festin2004}] or $(10^{-1}-10)$~Hz in the $R$ noise study in NbN \cite{Koushik2013}. The ability to obtain high-quality data at extremely low frequencies (long time scales) not only allows probing of the critical regime very close to $T_{BKT}$, where $\tau_{\xi}$ diverges, but also enables the extraction of reliable higher-order statistics, which provides information about correlations.

\section{Results and discussion}
\subsection{Resistance noise}

\begin{figure}
\includegraphics[width=8.5cm]{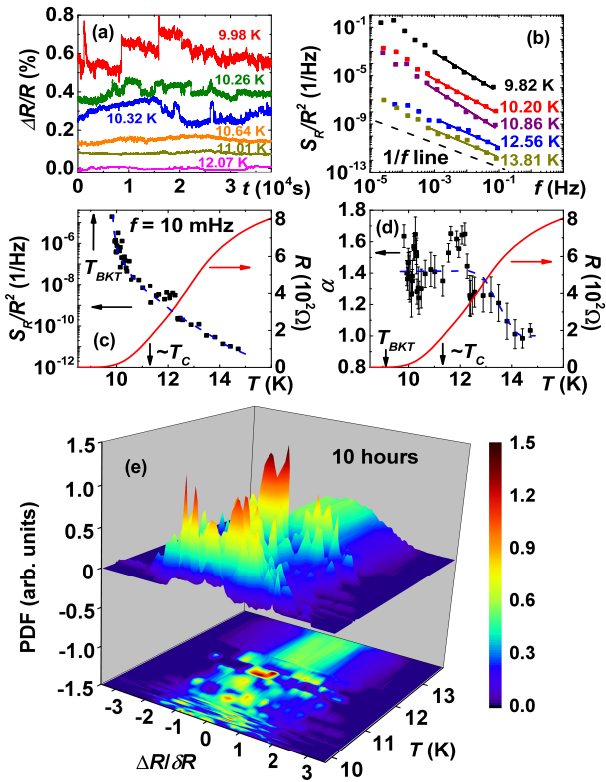}
\caption{$x=0.08$ LSCO film.  (a) $\Delta R(t)/R$ at several $T$.  All traces are shifted for clarity.  (b) The octave-averaged, normalized power spectra $S_{R}/R^{2}$ vs $f$ for several $T$. Solid lines are linear fits to $S_{R}/R^{2}\propto 1/f^{\alpha}$.  (c)  $S_{R}(f=10$~mHz)$/R^{2}$, determined 
from the fits in (a), vs $T$.  (d) $\alpha$ vs $T$.  In (c) and (d), solid lines show $R(T)$, and the arrows mark $T_{BKT}$ and $T_{c}$. The dashed lines guide the eye.  (e) PDF of the fluctuations at 9.8~$\leq T$(K)$\leq$~14.9 for 10-hour measurement times at each $T$.  For each time series, $\Delta R$ is normalized by the corresponding $\delta R=\langle [\Delta R(t)]^{2}\rangle ^{1/2}$, where $\langle\cdots\rangle$ denotes averaging over time, in order to make the change in the character of the PDF as a function of $T$ more apparent. \label{Fig2}}
\end{figure}

Figure~\ref{Fig2}(a) shows the typical time traces of the relative changes in resistance ${\Delta}R(t)/R$, or noise, at various $T$ for the $x=0.08$ film. 
%
%
As $T$ approaches $T_{BKT}$, the noise increases, and the system exhibits random fluctuations on many time scales, from rapid switching to slow changes over several hours.  Indeed, the corresponding power spectra (i.e. the Fourier transforms of a temporal autocorrelation function \cite{Hooge1969,Dutta1981,Weissman1988}) $S_{R}(f)/R^{2}$ [Fig.~\ref{Fig2}(b)] obey the well-known empirical law $S_{R}(f)/R^{2}\propto 1/f^{\alpha}$, reflecting a wide distribution of relaxation times.  In order to compare the noise magnitudes under different conditions, $S_{R}(f=10$~mHz)$/R^{2}$ is taken as the measure of noise, and its dependence on $T$ is shown in Fig.~\ref{Fig2}(c).  It is striking that, as $T$ decreases from $\sim 15$~K by only a few K, the noise increases by six orders of magnitude.  At the same time, $\alpha$ begins rising from $\sim 1.0$ before saturating at $\sim 1.4$ at $T\lesssim 12$~K [Fig.~\ref{Fig2}(d)].  This shift of the spectral weight towards lower $f$ indicates a dramatic slowing down of the dynamics as $T\rightarrow T_{BKT}^{+}$.  

The large values of $\alpha$ at low $T$ also reflect the non-Gaussianity of the noise, which is apparent already from the raw data [Fig.~\ref{Fig2}(a)].  Indeed, the analysis of the histograms or the probability density function (PDF) of the fluctuations reveals that the noise is Gaussian at $T\gtrsim 12$~K, but at lower $T$, the PDF  acquires a non-Gaussian, complex, multipeaked structure [Fig.~\ref{Fig2}(e)].  Most of the peaks result from the switching events observed in ${\Delta}R(t)/R$ [Fig.~\ref{Fig2}(a)], and the precise shape of the PDF obviously depends on the observation time.  These results indicate that very few states contribute to $R$ even for measurements as long as 10 hours [Fig.~\ref{Fig2}(e)] and that different states contribute to $R$ as a function of time.  Therefore, the system is nonergodic on experimental time scales \cite{Palmer}.  Similar out-of-equilibrium behavior was found in spin \cite{Weissman1993} and Coulomb glasses   \cite{Jaroszynski2002,Jaroszynski2004,Lin2012,Raicevic2008,Raicevic2011},  systems that wander collectively, with time, between many metastable states.  We note that here the observed change in the dynamics occurs near $T_c$, i.e. as the system crosses over from the high-$T$ regime of Aslamazov-Larkin Gaussian SC fluctuations to the BKT regime at lower $T$ \cite{Baity2016}.  

\subsection{Higher-order statistics}

The second spectrum $S_{2}(f_{2},f)$ \cite{Weissman1988}, a fourth-order statistic, is used to distinguish between a small number of independent single-rate processes with a distribution of rates, which may also give rise to high values of $\alpha$ \cite{Dutta1981,Weissman1988}, and correlated dynamics.  $S_{2}(f_{2},f)$, which is the power spectrum of the fluctuations of $S_{R}(f)/R^{2}$ with $t$, is white (independent of $f_2$) for independent fluctuators (Gaussian noise), and $S_{2}(f_{2},f)\propto 1/f_{2}^{1-\beta}$ for interacting ones \cite{Weissman1993,Dutta1981,Weissman1988,Weissman1992,Seidler1996a,Abkemeier1997}.  

\begin{figure}
\includegraphics[width=8.5cm]{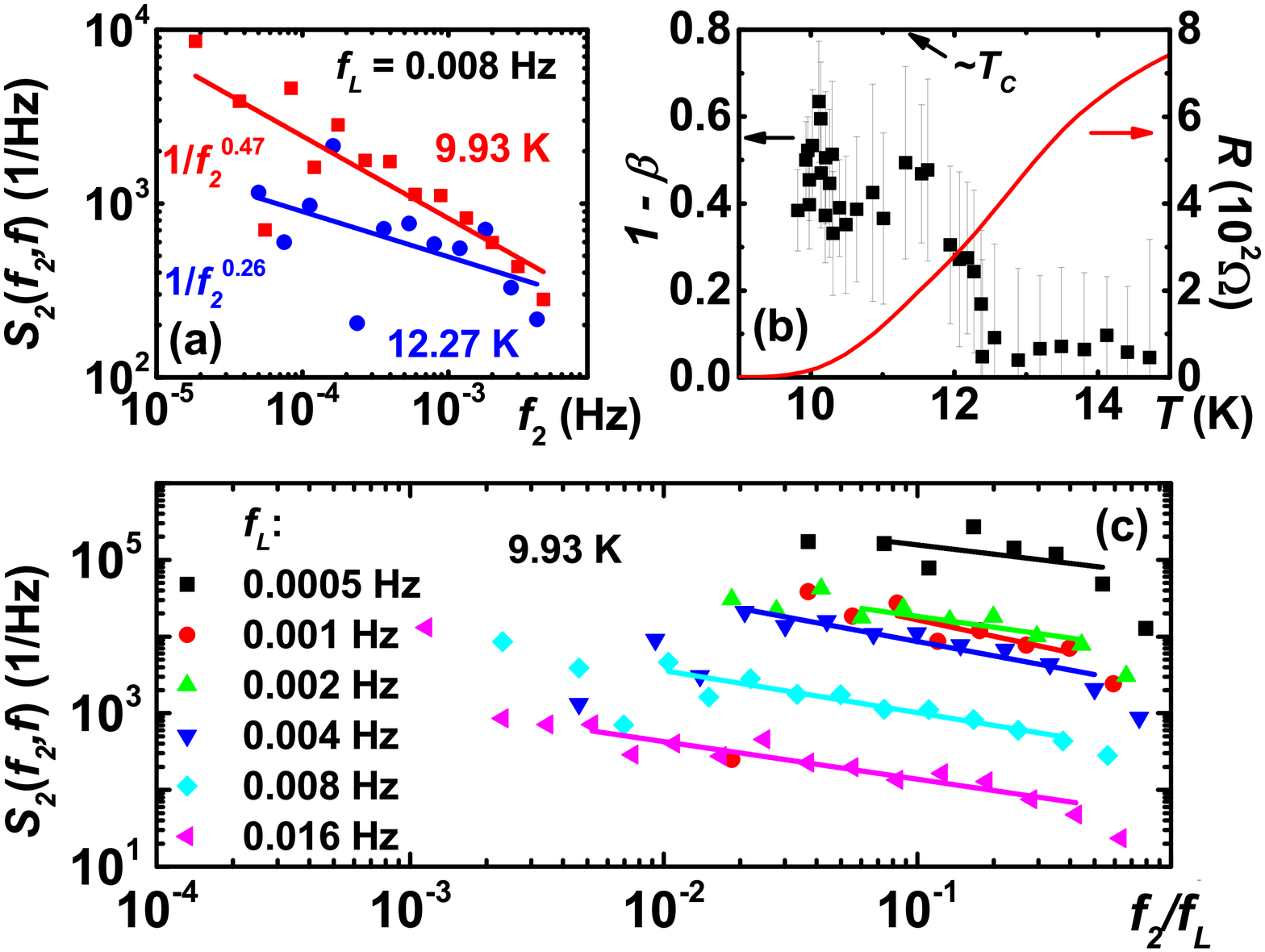}
\caption{$x=0.08$ LSCO film.  (a) The normalized second spectra $S_{2}(f_2,f)$, with the Gaussian background subtracted, for two $T$; $f_L= 8$~mHz.  Solid lines are fits to $S_2\propto 1/f_{2}^{(1-\beta)}$.  (b) $(1-\beta)$ vs $T$ for $f_L= 8$~mHz.  The same result was obtained for other $f_L$.  The solid line shows $R(T)$.  (c) $S_2$ for several $f=(f_L,2f_L)$ vs $f_2/f$ at $T=9.93$~K.  Solid lines are linear fits.  Qualitatively the same behavior is seen for $S_2$ vs $f_2$. \label{Fig3}}
\end{figure}

$S_2$ was calculated for a few octaves $f=(f_L,2f_L)$ at each $T$ [e.g.  Fig.~\ref{Fig3}(a)].
%
%
The results [Fig.~\ref{Fig3}(b)] show clearly an increase of $(1-\beta)$ from $\approx 0$ at $T\gtrsim 12$~K to nonwhite values at lower $T$, and a saturation at $\sim 0.5$ for $T\lesssim 11.6$~K ($\sim T_c$).  The onset of correlated dynamics as $T$ is reduced suggests that the probed time scales become shorter than $\tau_{\xi}$, i.e. probed length scales shorter than $\xi$.  Thus, the system appears to be critical. 

In order to explore the dynamics in the critical regime further, we compare $S_{2}(f_{2},f)$ for different $f$.  In Fig.~\ref{Fig3}(c), $S_{2}(f_{2},f)$ are plotted vs $f_2/f$, since spectra taken over a fixed time interval average the high-frequency data more than the low-frequency data.  In the studies of glasses, it was argued \cite{Weissman1993,Weissman1992}, albeit not calculated theoretically, that such $S_{2}(f_{2},f)$ should be scale invariant, or independent of $f$, in models in which the system wanders collectively between many metastable states related by a kinetic hierarchy.  Such scale-free behavior was indeed observed in metallic spin glasses \cite{Weissman1993,Weissman1992} and Coulomb glasses \cite{Bogdanovich2002,Jaroszynski2002,Jaroszynski2004,Lin2012}, both systems with long-range interactions.  In contrast, a decrease of $S_{2}(f_{2},f)$ with $f$ at constant $f_2/f$, similar to that in Fig.~\ref{Fig3}(c), was found in insulating LSCO \cite{Raicevic2008,Raicevic2011}, in the presence of an additional, effective short-range interaction.  In the study of glasses, such a decrease was proposed \cite{Weissman1993,Weissman1992} to be a feature of interacting droplet models \cite{Fisher1988b,Fisher1988a}.  However, it is not obvious that any such considerations would be applicable to the BKT critical regime, for which further theoretical work is clearly needed.  Nevertheless, an interesting question is whether the behavior found in Fig.~\ref{Fig3}(c) might reflect the presence of some other characteristic length scale, perhaps related to the inhomogeneity.

\subsection{Effects of disorder}

The previous study \cite{Baity2016} showed that these films are fairly clean, with a small degree of inhomogeneity leading to a larger smearing of the universal jump of $J_s$ for $x=0.07$ than for $x=0.08$. In particular, by performing a quantitative comparison of the data with theory, it was established that the width of the distribution of local $J_s$ values with respect to the most probable value, i.e., a measure of inhomogeneity, was ten times larger for $x=0.07$ than for the $x=0.08$ film (0.1 and 0.01, respectively) \cite{Baity2016}.  In addition, a larger normal-state resistivity or high-$T$ $R_{\square/\textrm{layer}}$ of the $x=0.07$ film (Fig.~\ref{Fig1}) indicates that it is more disordered than the $x=0.08$ sample.  For $x=0.07$, which is on the verge of a transition to an insulating state at lower $x$ \cite{Shi2013}, $R_{\square/\textrm{layer}}\lesssim h/e^2$ or $k_{F}l\gtrsim 1$ ($k_F$, Fermi wave vector; $l$, mean-free path).

\begin{figure}
\includegraphics[width=8.5cm]{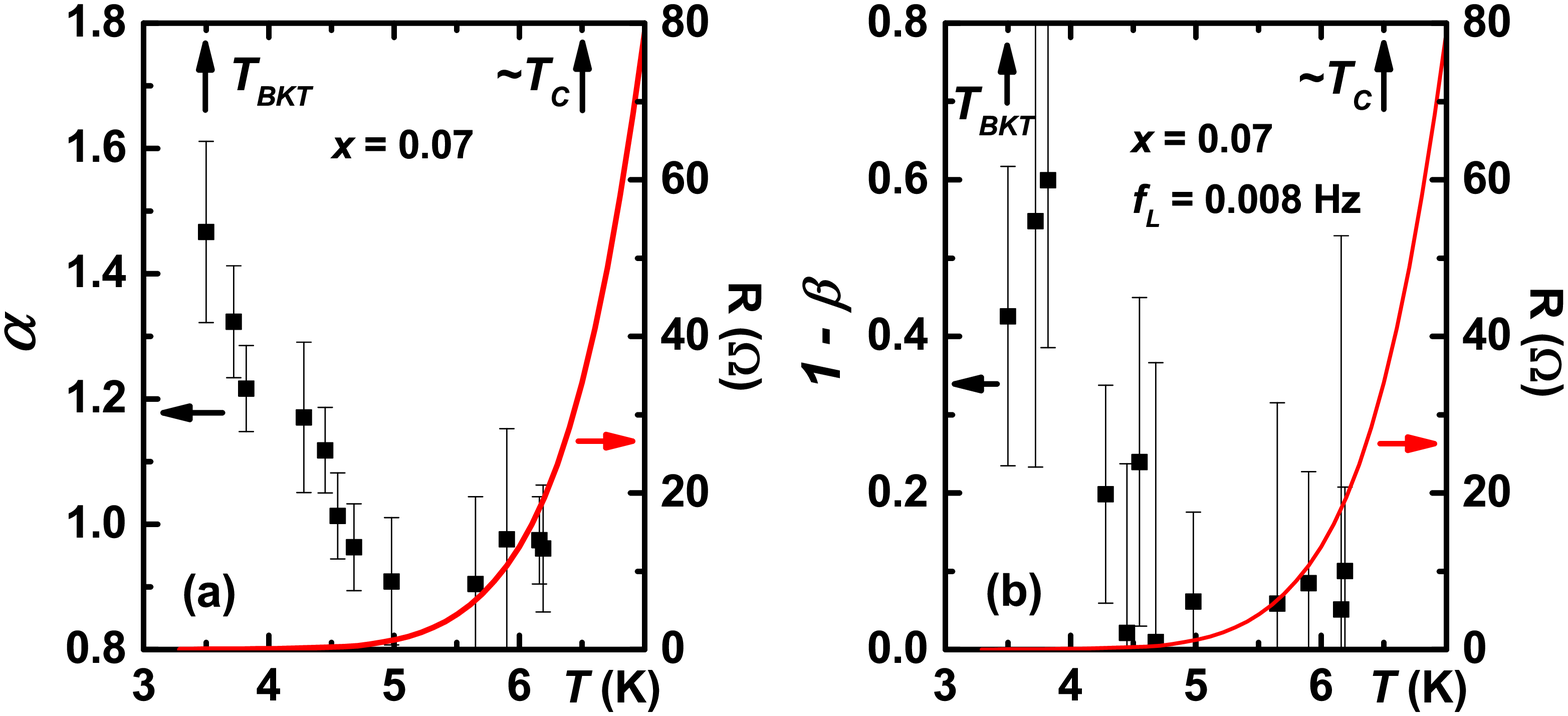}
\caption{$x=0.07$ LSCO film.  (a) $\alpha$ vs $T$. (b) $(1-\beta)$ vs $T$ for $f_L= 8$~mHz. Solid lines show $R(T)$. $T_{BKT}$ and $T_{C}$ are marked by arrows.
\label{Fig4}}
\end{figure}

The $R$ noise study reveals, however, that the onset of non-Gaussian noise is suppressed in the more disordered $x=0.07$ sample.  Indeed,  $\alpha$ rises from $\sim 1.0$ to $\sim 1.5$ as $T\rightarrow T_{BKT}^{+}$ also for $x=0.07$ [Fig.~\ref{Fig4}(a)], but  
%
%
in contrast to the $x=0.08$ film, the rise in $\alpha$ is observed only below $T\simeq4.5$~K~$\ll T_c$, with no visible saturation in the experimental $T$ range.  Furthermore, the analysis of the second spectrum reveals that $(1-\beta)$ becomes non-zero also below $T\sim 4.5$~K~$\ll T_c$, with an apparent saturation at $T< 4$~K [Fig. \ref{Fig4}(b)].  Therefore, for $x=0.07$, correlated dynamics is observed only at $T\ll T_c$, 
in contrast to $T\lesssim T_c$ for $x=0.08$.  The observed trend is precisely the opposite of what would be expected if the non-Gaussian noise was caused by disorder, and indicates that another mechanism, related to phase fluctuations, is responsible for the correlated behavior of the noise.

\begin{figure}
\includegraphics[width=8.5cm]{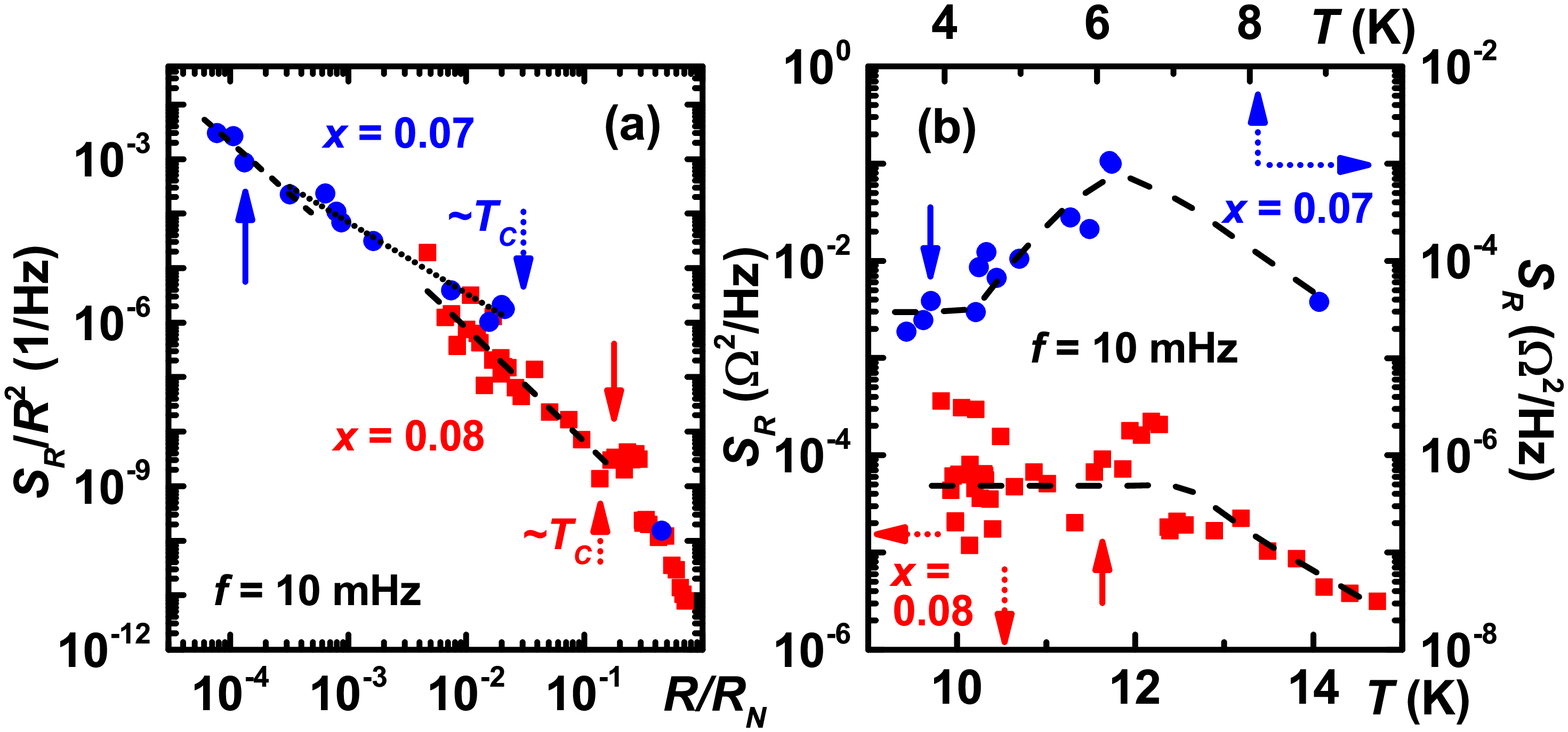}
\caption{(a) $S_{R}(f=10$~mHz)$/R^{2}$ vs $R/R_{N}$ for $x=0.07$ (blue dots) and $x=0.08$ (red squares) samples, respectively.  $R_{N}$ is chosen near the maximum of $R(T)$, where SC fluctuations are negligible \cite{Baity2016}: $R_{N}\equiv R(T=22$~K) and $R_{N}\equiv R(T=28$~K) for $x=0.07$ and 0.08, respectively.  Dashed lines are fits to $S_{R}/R^{2}\propto R^{-s}$ in the correlated regime, with slopes $s=(1.9\pm 0.3)$ and $s=(2.1\pm 0.2)$ for $x=0.07$ and 0.08, respectively.  A power-law fit (dotted line) for $x=0.07$ above the correlated regime, but at $T<T_c$, yields $s=(1.31\pm 0.09)$, and $s=(5.2\pm 0.5)$ for $x=0.08$ at $T>T_c$.  (b) $S_R(f=10$~mHz) vs $T$ for the two samples, as shown.  Dashed lines guide the eye.  In (a) and (b), solid arrows mark the onset of the low-$T$ saturation of $(1-\beta)$.
\label{Fig5}}
\end{figure}

The above conclusion is further supported by comparing the noise magnitudes vs $R/R_N$ for the two samples ($R_N$ is the normal state resistance) [Fig.~\ref{Fig5}(a)].  
%
%
For both samples, $S_{R}(f)/R^{2}\propto R^{-s}$ for $T<T_c$ [Fig.~\ref{Fig5}(a)], such that $s=2$ in the regime of correlated dynamics.  On the other hand, $s\approx 1.31$ for $x=0.07$ at $T<T_c$, but \textit{above} the correlated regime.  A power-law dependence of the noise magnitude on $R$ above the SC transition was found also in films of other cuprates \cite{Kiss1994}, but at a much higher doping and in the absence of the BKT regime.  The exponent $s=1.54$ was attributed to percolation \cite{Kiss1994,Kiss1993}, and so was $s\sim 1$ found in NbN films above the BKT transition \cite{Koushik2013}.  Hence, the very different value of $s$ found in the regime of correlated noise in Fig.~\ref{Fig5}(a) is consistent with the conclusion that it is not caused by disorder.  In fact, since the power spectrum $S_{R}(f)$ of resistance fluctuations $\Delta R(t)$ does not depend on $T$ in the correlated regime [Fig.~\ref{Fig5}(b)], then $S_{R}(f)/R^{2}\propto R^{-2}$, i.e. $s=2$.  Furthermore, the huge increase of the noise magnitude $S_R(f)/R^2$ as $T\rightarrow T_{BKT}$ [Fig.~\ref{Fig2}(c)] is then due to the exponential decrease of $R(T)$ \cite{Baity2016}, i.e. the exponential divergence of $\xi(T)$. 

We note that also the $T$ dependence of the correlated noise, shown in Figs.~\ref{Fig5}(b) and \ref{Fig2}(c), is different from that in ultrathin NbN films above the BKT transition \cite{Koushik2013}.  In fact, even the exponent $\alpha\lesssim1.5$ [Figs.~\ref{Fig2}(d) and \ref{Fig4}(a)] is quite different from the uncommonly large \cite{Weissman1993,Dutta1981,Weissman1988} $\alpha\sim 2-4$ in NbN  \cite{Koushik2013}, consistent with our conclusion that the correlated noise in LSCO reflects the intrinsic BKT dynamics.  For completeness, we also note that $T$ dependence of the correlated noise in $x=0.08$ and $x=0.07$ LSCO [Figs.~\ref{Fig5}(b) and \ref{Fig2}(c)] is different from that above the insulating  ground state in lightly doped ($x=0.03$) LSCO \cite{Raicevic2008,Raicevic2011}.

Outside of the correlated regime, $S_{R}(f)$ does depend on $T$ in both films [Fig.~\ref{Fig5}(b)].  In particular, for $T>T_c$, $S_{R}(f)$ decreases with increasing $T$ in both samples, but understanding of this regime is beyond the scope of this work.  Likewise, for $x=0.07$ at $T<T_c$ above the correlated regime, the $T$ dependence of $S_{R}(f)$ and the rather different value of $s$ could be related to disorder, but a detailed investigation of the effects of disorder on the BKT regime will be needed to understand this behavior.  

\section{Conclusions}

We have combined low-frequency resistance noise spectroscopy, an uncommon but powerful technique, with unconventional superconductors to probe the correlated dynamics near a BKT transition.  Several different noise statistics have been analyzed, from the full probability distribution of the fluctuations to the first spectrum (a second-order statistic) and second spectrum (a fourth-order statistic).  The results, which were obtained in highly underdoped thick films of La$_{2-x}$Sr$_{x}$CuO$_{4}$, demonstrate slowing down of the dynamics resulting from the exponential divergence of $\xi$ near a BKT transition.  In the same regime, where phase fluctuations dominate, we find that the dynamics is correlated and nonergodic, strongly suggesting that the experiment probes the low-$T$ critical phase.  Furthermore, we have established detailed properties of the noise [e.g. Figs.~\ref{Fig2}(d) and \ref{Fig3}(c)], which are important for developing proper theoretical understanding of the dynamics near the BKT transition.  Finally, we have determined that the correlated dynamics in LSCO is suppressed by disorder.  Indeed, several noise characteristics exhibit \textit{qualitatively} different behavior in LSCO than in NbN, in which disorder was found to have a key effect on the noise properties even in the correlated regime \cite{Koushik2013}.

It is interesting to speculate whether the correlated dynamics observed in LSCO reflects the ergodicity-breaking character of the BKT transition \cite{Faulkner2015}, or it might be related to the falling out of equilibrium upon cooling to near $T_{BKT}$ \cite{Jelic2011}.  Our work offers new insights into the dynamics of topological defects across thermal phase transitions, but further experimental and theoretical studies are needed to determine whether the observed features are common to other layered superconductors and to enable a more direct comparison to theory.
\vspace{12pt}

\begin{acknowledgments}
We thank A. T. Bollinger and I. Bo\v{z}ovi\'{c} for the samples. This work was supported by NSF grant No. DMR-1307075 and the National High Magnetic Field Laboratory through the NSF Cooperative Agreement No. DMR-1157490 and the State of Florida.
\end{acknowledgments}


\end{document}